\documentclass[final,5p,times]{elsarticle}

\usepackage{amssymb}

\usepackage{amsmath}
\usepackage{graphicx}
\usepackage{times}
\usepackage{caption}
\usepackage{subcaption}
\usepackage{flushend}
\usepackage[colorlinks = true,
	linkcolor = red,
	urlcolor = red,
	citecolor = red,
	anchorcolor = red]{hyperref}
\usepackage{verbatim}

\usepackage[nodots]{numcompress}

\usepackage{lineno}

\journal{Sensors}

\begin{document}
\begin{frontmatter}

\title{Full-Body Locomotion Reconstruction of Virtual Characters Using a Single IMU}

\author{Christos Mousas}

\address{Department of Computer Science\\ Southern Illinois University, Carbondale, IL 62901, USA\\
\url{christos@cs.siu.edu}
}

\begin{abstract}
This paper presents a method of reconstructing full-body locomotion sequences for virtual characters in real-time, using data from a single inertial measurement unit (IMU). This process can be characterized by its difficulty because of the need to reconstruct a high number of degrees of freedom (DOFs) from a very low number of DOFs. To solve such a complex problem, the presented method is divided into several steps. The user's full-body locomotion and the IMU's data are recorded simultaneously. Then, the data is preprocessed in such a way that would be handled more efficiently. By developing a hierarchical multivariate hidden Markov model with reactive interpolation functionality the system learns the structure of the motion sequences. Specifically, the phases of the locomotion sequence are assigned in the higher hierarchical level, and the frame structure of the motion sequences are assigned at the lower hierarchical level. During the runtime of the method, the forward algorithm is used for reconstructing the full-body motion of a virtual character. Firstly, the method predicts the phase where the input motion belongs (higher hierarchical level). Secondly, the method predicts the closest trajectories and their progression and interpolates the most probable of them to reconstruct the virtual character's full-body motion (lower hierarchical level). Evaluating the proposed method shows that it works on reasonable framerates and minimizes the reconstruction errors compared with previous approaches.
\end{abstract}


\end{frontmatter}


\section{Introduction}
\label{sec1}

\begin{figure*}
\centering
\includegraphics[width=1\textwidth]{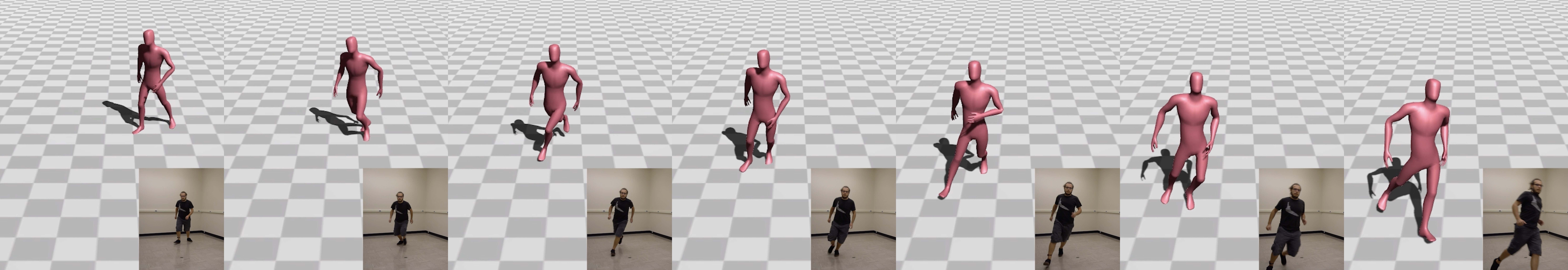}
\caption{A reconstructed running motion when the user follows a curved path.}
\label{fig1}
\end{figure*}

In modern computer animation production, motion capture solutions simplify animators' lives. Motion capture systems not only quickly capture human motion but also its naturalness. However, most motion capture solutions are expensive (except a few affordable systems). Most of them also require special equipment and extensive preparation time to place the necessary markers on a performer's body and calibrate the system. A final disadvantage of most high-quality systems is that they are almost always limited to indoor studios, which is especially true for optical systems.

Recently, extensive research has focused on reconstructing human motion from a low-cost inertial measurement unit (IMU). One research direction has aimed to reduce the number of sensors or markers required. Most motion reconstruction methods such as \cite{ref26}\cite{ref50} use six sensors (hands, feet, head, and root) and a motion database containing sample poses of a virtual character. Thus, using the signals from the sensors, either kinematic solutions or data-driven methodologies for reconstructing full-body motion using mainly statistical motion models have been developed.

Considering the prohibitive cost of acquiring a high-quality motion capture system and the difficulties associated with reconstructing the locomotion sequences for large numbers of people (e.g., capturing the locomotion sequences of a group of pedestrians to analyze their reactions and interactions in order to compose virtual crowds), affordable solutions that provide such functionality are highly desirable. Based on this aim, this paper presents a method of reconstructing the full-body locomotion behaviors of a virtual character by using data provided from a single sensor. It sound be noted that nowadays, a variety of devices, such as smartphones and watches, have an embedded accelerometer and gyroscope. 

To reconstruct motion using a single IMU, a probabilistic model based on prerecorded human motion data was developed. First, a multivariate hidden Markov model (HMM) was used to map the performer's retrieved feature vectors (captured from the full-body motion capture device) to those retrieved from the single IMU. The multivariate HMM is extended to handle hierarchical levels of motion. Specifically, the model is trained to recognize the phases of the locomotion that a user performs at its higher hierarchical level, and to reconstruct the virtual character's full-body motion at its lower hierarchical level through interpolating the closest trajectories. Reconstructing full-body motion sequences pose-by-pose can produce ambiguity and temporal incoherence. Thus, a moving window and the forward algorithm were used during the application's runtime to determine the phase where the user's input motion belonged and then to reconstruct the virtual character's full-body locomotion.

This research makes several important contributions. First, it proposes a novel way of reconstructing a virtual character's full-body locomotion by using a single IMU. Second, it shows how the proposed approach can reconstruct different locomotion behaviors (walking, running, etc.) based on a single training process. For example, Eom et al. \cite{ref12} reconstructed each locomotion separately (the model is trained to recognize a character's particular locomotion behavior). Third, the proposed method makes fewer reconstruction errors than those of previous solutions. Finally, the proposed hierarchical multivariate HMM with the reactive interpolation functionality can also be considered a novelty. The presented probabilistic model for a virtual character's locomotion reconstruction extends a previously proposed hierarchical HMM for character animation \cite{ref49} by introducing the reactive model interpolation functionality. The advantages of the proposed probabilistic model may also benefit a variety of applications related to computer animation and interactive character control.

This paper is organized as follows. Section \ref{sec2} discusses the work related to the proposed approach. Section \ref{sec3} describes the methodology used in the proposed method. Section \ref{sec4} explains the preprocessing steps. Section \ref{sec5} covers the proposed HMM that was developed to handle this complex locomotion reconstruction problem. Section \ref{sec6} presents the motion reconstruction process based on the forward algorithm. The evaluations that were conducted are described in Section \ref{sec7}. Finally, Section \ref{sec8} concludes with a discussion of the methodology's limitations.

\begin{figure*}
\centering
\includegraphics[width=1\textwidth]{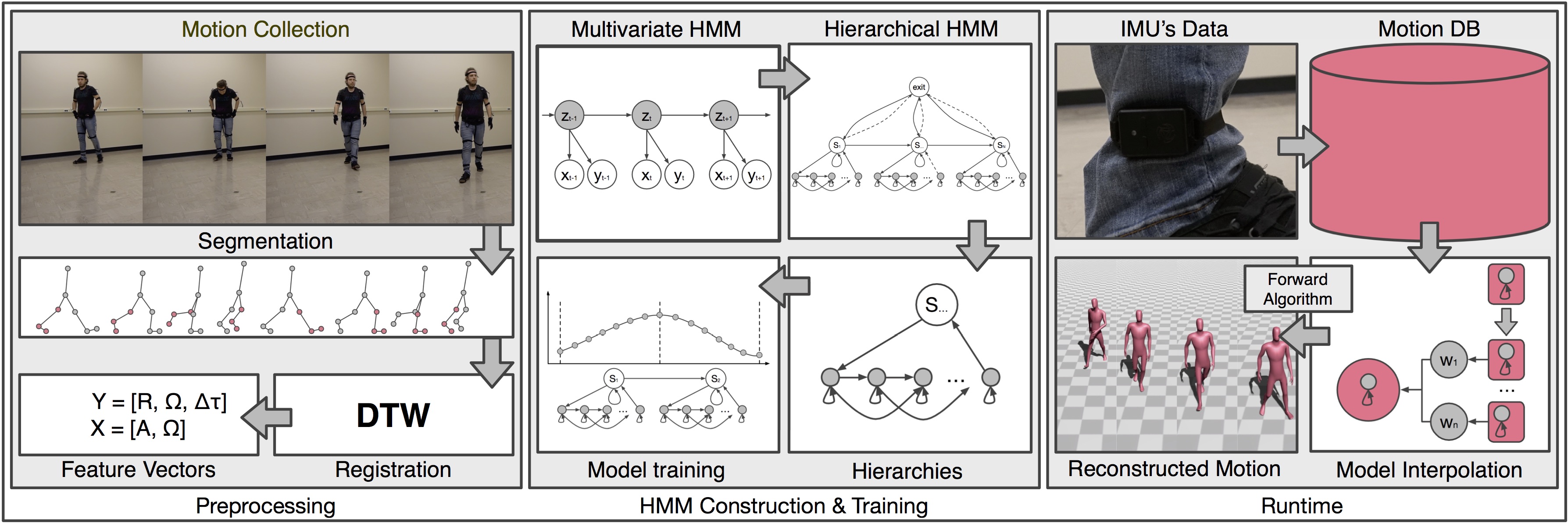}
\caption{Overview of the methodology.}
\label{fig2}
\end{figure*}

\section{Related Work}
\label{sec2}
Motion capture solutions are used extensively in a variety of entertainment and computer animation-related areas and applications. In recent years, various solutions have been developed, with high-quality results. Motion capture systems can be divided into different categories based on the technologies that they use. The main categories are optical, mechanical, magnetic, and inertial. Each system type has its own advantages and disadvantages regarding the accuracy of the captured data, the capability to capture a particular volume, and the operation effort required. A more detailed explanation of how motion capture systems work can be found in \cite{ref14}\cite{ref15}\cite{ref59}.

This paper proposes a method of reconstructing a character's full-body motion by using a reduced number of data inputs in conjunction with the ability of statistically analyzing sample motion data. This approach applies to performance animation and motion reconstruction. In performance animation techniques (known as ``computer puppetry" \cite{ref17}\cite{ref18}), performers can manipulate the body parts of a virtual character by using kinematic solutions \cite{ref19}\cite{ref20}, the system can recognize the performer's action (activity recognition) and display the motion from a database \cite{ref21}\cite{ref22}\cite{ref23}\cite{ref24}\cite{ref25}\cite{ref27}, or the system can synthesize a new motion sequence by using the existing motion data in a database (motion reconstruction) \cite{ref13}\cite{ref26}\cite{ref28}\cite{ref29}\cite{ref30}. In these three approaches, the input signals provided by the motion capture device are used as the parameters for the motion control or reconstruction process. To animate the virtual character, methodologies that use accelerometers \cite{ref22}\cite{ref13}\cite{ref31} or optical motion capture devices \cite{ref32}\cite{ref33} provide the desired control parameters for the system. Recent research has focused on the ability to synthesize natural-looking motion sequences while using a reduced number of input data. Hence, methods that use six \cite{ref26} or two \cite{ref11} inputs (markers or sensors) or even just one \cite{ref12} are capable of reconstructing a character's motion in real-time. These kinds of methods are commonly based on the construction and the use of statistical analysis of prerecorded human motion data \cite{ref26}. Such methods map the reduced input parameters to a database of postures to find and synthesize new motion sequences that match the input constraints. A few previously proposed methodologies that use a reduced number of input signals to reconstruct the virtual character's full-body motion are described in the following paragraphs. 

A statistical analysis and synthesis method of complex human motions were introduced by Li et al. \cite{ref51}. This method provided results that were statistically similar to the original motion captured data. Based on a real-time inverse kinematic solver, Badler et al. \cite{ref36} were able to control a virtual character by using data from only four magnetic sensors. In Semwal et al.'s study \cite{ref37}, eight magnetic sensors in conjunction with an analytical mapping function were able to control a virtual character's motion. A foot pressure sensor was used by Yin and Pai \cite{ref24} to reconstruct the poses of a virtual character. Chai and Hodgins \cite{ref3} developed a real-time performance-driven character control interface that used cameras to capture the markers attached to a performer. Their methodology was based on a local motion model, which was utilized with numerous prerecorded motion capture data. Pons-Moll et al. \cite{ref38} proposed a hybrid motion-tracking approach that combined inputs from video data with a reduced number of inertial sensors. 

By extending the methodology developed by Cai and Hodgins \cite{ref3}, Tautges et al. \cite{ref39} experimented on human motion tracking while changing the number of the sensors. Tautges et al. \cite{ref13} also developed a method of reconstructing a character's full-body motion by using only four accelerometers. Each one was attached to the end effectors of a performer. In their study, the authors experimented with a variety of scenarios. They tested their system's ability to efficiently reconstruct the virtual character's motion by changing the number of sensors, as well as by positioning the sensor on different body parts. Their approach demonstrated the possibility to reconstruct various smooth motions by only using four sensors, which is the optimal number according to their results.

In some cases, human motion should be reconstructed with a reduced number of degrees of freedom (DOFs). For example, stepping and walking activities can be easily reconstructed using only six DOFs, as proposed by Oore et al. \cite{ref40}. By utilizing a simple game controller, Shiratori and Hodgins \cite{ref31} synthesized virtual characters' physically based locomotion sequences. Finally, von Marcard et al. \cite{ref50} recently proposed a method of reconstructing complex full-body motion in the wild by using six IMUs. This method has the advantage of taking into account anthropometric constraints in conjunction with a joint optimization framework to fit the model.

The approach presented in this paper is similar to previous methods in terms of reconstructing the full-body locomotion of a virtual character. Compared with Slyper and Hodgins' \cite{ref22} method, the present one not only searches in a motion database for the closest motion but also interpolates between them to reconstruct a new motion that follows the user's input trajectory. Smoother results with fewer reconstruction errors were achieved compared with Tautges et al.'s method \cite{ref13}. The closest solution to the approach presented here is the one proposed by Eom et al. \cite{ref12}. The major disadvantage of Eom et al.'s method is that it can only reconstruct a single locomotion behavior at a time and cannot reconstruct motions such as running on a curved path (see Figure \ref{fig1}) or even continuous locomotion with multiple behaviors (see accompanying video). In contrast, the proposed method can reconstruct any locomotion type (e.g., walking, running, jumping, hopping, and side stepping) in any row with any variation (e.g., turning during walking or turning during running). Therefore, the main advantage of the presented statistical model is its ability to reconstruct long locomotion sequences with different behaviors in real-time by using input data provided by a single sensor.

\section{Methodology}
\label{sec3}
The methodology is divided into three phases. The first phase meets several preprocessing requirements. The motion collection part simultaneously captures the performer's full-body motion and the IMU's data. The motion segmentation part segments the full-body motion and IMU data into meaningful phases of locomotion. The motion registration process registers the segments in a way that helps the statistical model reconstruct a motion sequence more efficiently. The final preprocessing step is the composition of the feature vectors that describe each motion segment. In the second phase, the proposed hierarchical multivariate HMM with the reactive interpolation functionality is constructed and trained. The constructed hierarchical HMM with reactive interpolations functionality is responsible for handling the motion data. Specifically, the higher state of the HMM encodes the segment phases of the locomotion sequences and the lower level of the HMM encodes the time index of the segment phases. This structure allows the prediction of the locomotion phase and later the prediction of the progress of the motion itself. The final phase is the application's runtime, where the system reconstructs the virtual character's motion based on the user's input performance (data retrieved from the IMU attached to the user). During the runtime, the reactive interpolation functionality was implemented to blend the nearest neighbor segments, resulting the reconstructed motion to more precisely follows the performance of the user. These phases are summarized in Figure \ref{fig2} and presented in the following sections.

\section{Preprocessing}
\label{sec4}
This section presents the preprocessing parts of the proposed method.

\begin{figure}[ht]
\centering
\includegraphics[width=1\columnwidth]{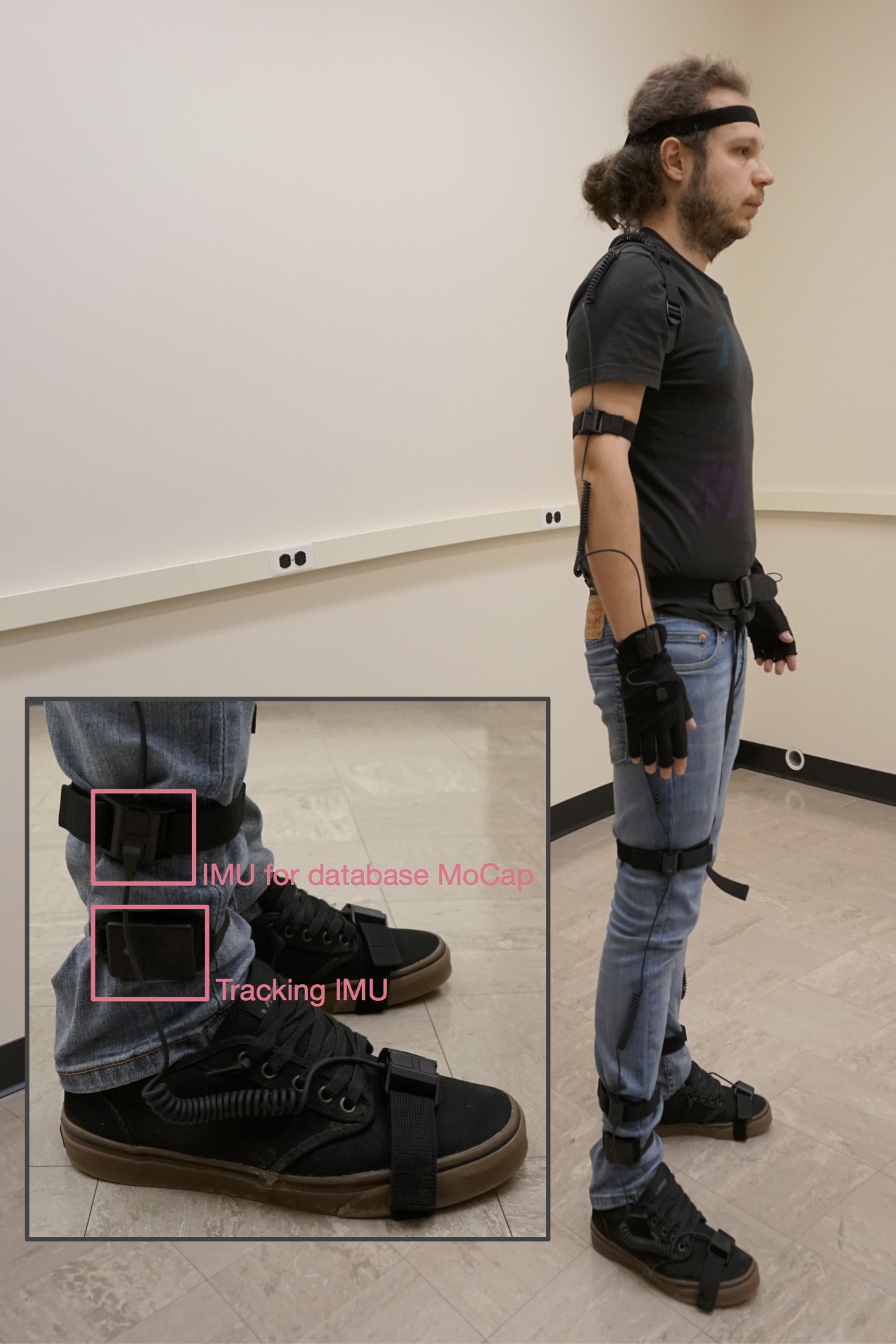}
\caption{A performer wearing the motion capture suit with the IMU attached to his leg.}
\label{fig3}
\end{figure}

\subsection{Capturing Motion Data}
\label{sec41}
During the first preprocessing step, the performer's full-body motion and the IMU's data were captured by asking the performer to wear the appropriate capture suit, with the required IMU attached to his leg (see Figure \ref{fig3}). The same IMU was also used in the real-time reconstruction process. In both cases, after capturing the motion data, the sampling rate was reduced to 30 frames per second. The associated software development kits (SDKs) of the motion capture system and the sensor were used in this implementation. A wrapper software was developed based on the two SDKs that simultaneously captured the input data from the full-body suit and the single sensor, which was the aim of this step. If the data are not captured simultaneously, an alignment method can easily address the alignment problem.

\subsection{Motion Segmentation}
\label{sec42}
\begin{figure*}[htb]
\centering
\includegraphics[width=1\textwidth]{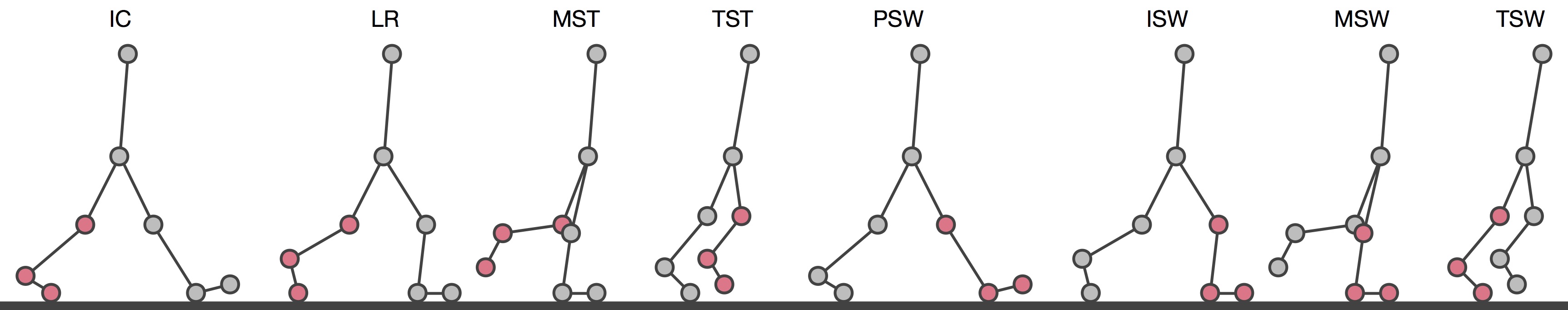}
\caption{Eight phases of the human gait cycle.}
\label{fig4}
\end{figure*}

The captured motion data was segmented into discrete and meaningful locomotion phases. Of the various ways of segmenting human locomotion, most are based on the phases of the locomotion (e.g., in single- and double-limb stances). This segmentation method did not differ from previous solutions because it also segmented the motion data according to its phases. However, as the problem was quite complex, the motion data was segmented into shorter phases (compared with \cite{ref12} and \cite{ref11}). Based on the studies of Ayyappa \cite{ref1}, Marks \cite{ref35}, and Loudon et al. \cite{ref42}, the presented segmentation process considered the eight phases characterizing a single cycle of the human gait: initial contact (IC), loading response (LR), mid-stance (MST), terminal stance (TST), pre-swing (PSW), initial swing (ISW), mid-swing (MSW), and terminal swing (TSW) (see Figure \ref{fig4}). 

To achieve this segmentation process, a simple foot contact with the ground method was combined with a crossing event (CE) at the ankle \cite{ref34}. The contact with the ground method reported whether or not the heel or the toe made contact with the ground. The CE checked whether one ankle (either right or left) was behind or in front of the other foot (either right or left), such as whether the right ankle was behind the left foot. Table \ref{tab1} describes the details of the segment phases and how the locomotion sequences were segmented. This combination made it possible to segment the motion data into eight phases. Shorter segments (compared with \cite{ref11} and \cite{ref12}) were used because the constructed model required the ability to exit a segment state quickly (and consequently, to enter another quickly). This approach allowed the system to reconstruct different locomotion sequences almost instantaneously. This ability would be especially useful in case users wanted to change which motion type they were performing. Finally, while this process was used to segment locomotion sequences, such as walking and running, the results showed that the presented method could also handle and reconstruct other behaviors, such as jumping, hopping, side stepping. In such cases, the contact with the ground approach was combined with the speed (minimum and maximum) approach to segment the motion data, similar to the segmentation used by Eom et al. \cite{ref12}.

\begin{table}
\centering
\caption{Details of the segment phases and how locomotion sequences were segmented. The segmentation process uses contact with the ground information of the foot parts and a crossing event (LCE for left and RCE for right foot) if the ankle is in front of or behind the foot and are abbreviated as follows: right toe contact (RT), right heel contact (RH), left toe contact (LT), left heel contact (LH), right crossing event (RCE), and left crossing event (LCE).}
\begin{tabular}{| p{2.9cm} | p{2.4cm} | p{2.4cm} |}
\hline
\textbf{Phase} 		& \textbf{Start Stance Conditions} 	& \textbf{End Stance Condition}\\
\hline
\hline
From IC to LR 		& RH \& LT 					& RH \& RT \& !LT\\
\hline
From LR to MST 		& RH \& RT \& !LT 				& RH \& RT \& LCE\\
\hline
From MST to TST 		& RH \& RT \& LCE 				& RT \& !LH \& !LT\\
\hline
From TST to PSW 	& RT \& !LH \& !LT 				& RT \& LH\\
\hline
From PSW to ISW 	& RT \& LH 					& !RT \& LH \& LT \\
\hline
From ISW to MSW 	& !RT \& LH \& LT 				& RCE \& LH \& LT\\
\hline
From MSW to TSW 	& RCE \& LH \& LT 				& !RH \& !RT \& LT\\
\hline
From TSW to IC 		& !RH \& !RT \& LT 				& RH \& LT \\
\hline
\end{tabular}
\label{tab1}
\end{table}

\subsection{Full-Body and IMU Data Registration}
\label{sec43}
After capturing and segmenting the datasets (full-body motion and IMU), both the full-body motion and the IMU data had to be registered. Notably, the trajectories from both the full-body and IMU (trajectories are retrieved from the angular velocity of the IMU) sequences were registered in the same way. However, the full-body motion sequences were mainly registered. By maintaining the parameters of the registration process it was possible to apply the parameters to the captured trajectories of the IMU in order to follow the same registration as that of the full-body motion data. Each motion segment that corresponded to a particular phase of the locomotion sequence was registered against each other. Specifically, one of the segments was picked as reference (the corresponding IMU's segment was also picked as a reference) and was used to register the rest of motion segments via the appropriate time-warping function. The motion sequences were registered to one another based on their translation and rotation by decomposing each pose of a segment from its translation on the ground plane and rotating its hips around the vertical axis, similar to Kovar and Gleicher's proposed method \cite{ref5}. Then, a dynamic time-warping technique was used to register all of the motion segments that belonged to a particular locomotion phase. Next, each motion segment was warped to the reference motion segment, and the associated timing was estimated by using the necessary time warping. Later, the corresponding time-warping functions were used to handle the trajectories retrieved from the sensor, ensuring that the correspondences between the data were maintained. This registration process provided the ability to decouple the motion segments belonging to a particular locomotion phase, making them suitable for use in the motion reconstruction processing.

\subsection{Motion Features}
\label{sec44}

The presented methodology was developed to predict and reconstruct a character's full-body motion from a single sensor. Computing a number of motion features for both captured data was required to provide this functionality. Later, these feature vectors were mapped to one another. From now on, $X$ denotes the feature vector of the full-body motion sequence, and $Y$ signifies the feature vector of the IMU's data. Moreover, $x_t$ and $y_t$ represent the corresponding $X$ and $Y$ feature vectors at time $t$.

For the full-body motion, three features were considered: the rotation of the joints $r_t$, the angular velocity of the joints $\omega_t$, and the root position difference $\Delta\tau$ between $t-1$ and $t$, which is defined by $\Delta\tau_{t} = \tau_{t-1}-\tau_t$. Thus, the feature vector is represented as $X=[R,\Omega,\Delta\tau]$, where at time $t$ is defined as $x_t=[r_t, \omega_t, \Delta\tau_t ]$, where $x_t \in \mathbb{R}^{d_x}$. The rotation of the joints is represented as $r_t=[r_t (1),...,r_t (N)]$, and the angular velocity is represented as $\omega_t=[\omega_t (1),...,\omega_t (N)]$. In both cases, $N$ denotes the total number of joints, and $t=1,...,T$ represents the complete number of frames of the motion data. It should be noted that the root position difference feature, $\Delta\tau$, is quite important for updating the global position of the character's root. For the sparse IMU, it is possible to compute the acceleration $e_t$ that captures meters per second squared ($m/sec^{2}$) and the angular velocity $\omega_t$, provided by the gyroscope that captures $rad/s$. Therefore, the feature vector is represented as $Y=[E,\Omega]$, where at time $t$ is represented as $y_t=[e_t, \omega_t]$, where $y_t \in \mathbb{R}^{d_y}$. It should be noted that $T$ is the same for both datasets since they were captured simultaneously. In cases where the motion data was captured separately, an alignment process and a resampling might be required.

\section{Hierarchical Multivariate HMM with Reactive Interpolations}
\label{sec5}

\begin{figure*}[htb]
\centering
\includegraphics[width=1\textwidth]{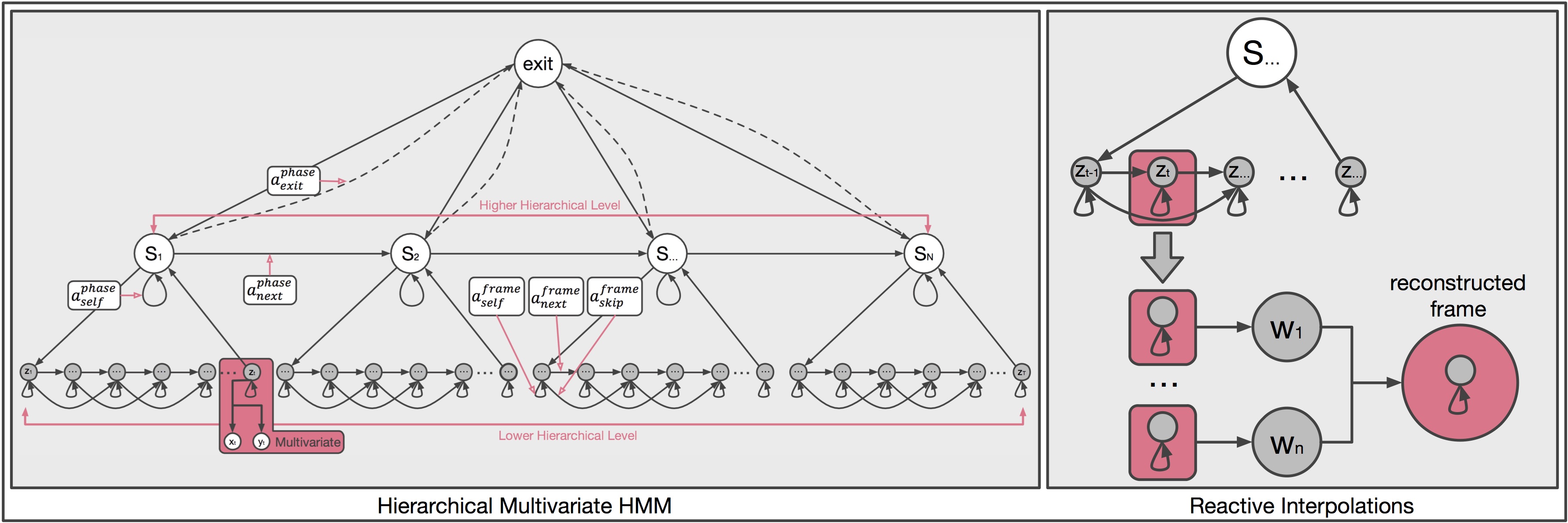}
\caption{Representation of the constructed hierarchical multivariate HMM with the reactive interpolations.}
\label{fig5}
\end{figure*}

The following subsections describe the construction of the hierarchical multivariate HMM with the reactive interpolation functionality. Specifically, the subsections present the multivariate HMM process that mapped the captured data, the way that the hierarchical HMM was constructed and its training process, as well as how the reactive model interpolation was achieved. Figure \ref{fig5} illustrates a graphical representation of the proposed hierarchical multivariate HMM with the reactive interpolations functionality. The model encoded the segmented phase of the locomotion and the frames of the motion at the higher and the lower hierarchical levels, respectively. The lower level was also responsible for combining the closest frames to reconstruct the full-body motion of a character from the IMU's data. From now on, the parameters $\lambda$ of the HMM are defined as $\lambda = \{ a_{ij},\pi_i,b_i \} $, where $a_{ij}$ (state transition matrix) denotes the probability of making a horizontal transition from the $i-th$ state to the $j-th$ state, $\pi_i$ (prior vector) represents the initial distribution vector over the substates, and finally $b_i$ (observation probability distribution) indicates the probability of the production state.

The rest of this section is organized as follows. Sub-section \ref{sec51} presents the multivariate HMM (regression process) that was implemented to handle the input data from the IMU and the full-body motion of the virtual character. Sub-section  \ref{sec52} describes the entire hierarchical structure of the HMM that was developed to handle the phases of the locomotion (higher level) and the progress of the motion segments (lower level). Finally, Sub-section \ref{sec53} presents the reactive interpolation functionality that was implemented to blend the nearest neighbor segments in order to provide a full-body motion that more precisely follows the performance of the user.

\subsection{Multivariate HMM Mapping}
\label{sec51}

\begin{figure}[htb]
\centering
\includegraphics[width=1\columnwidth]{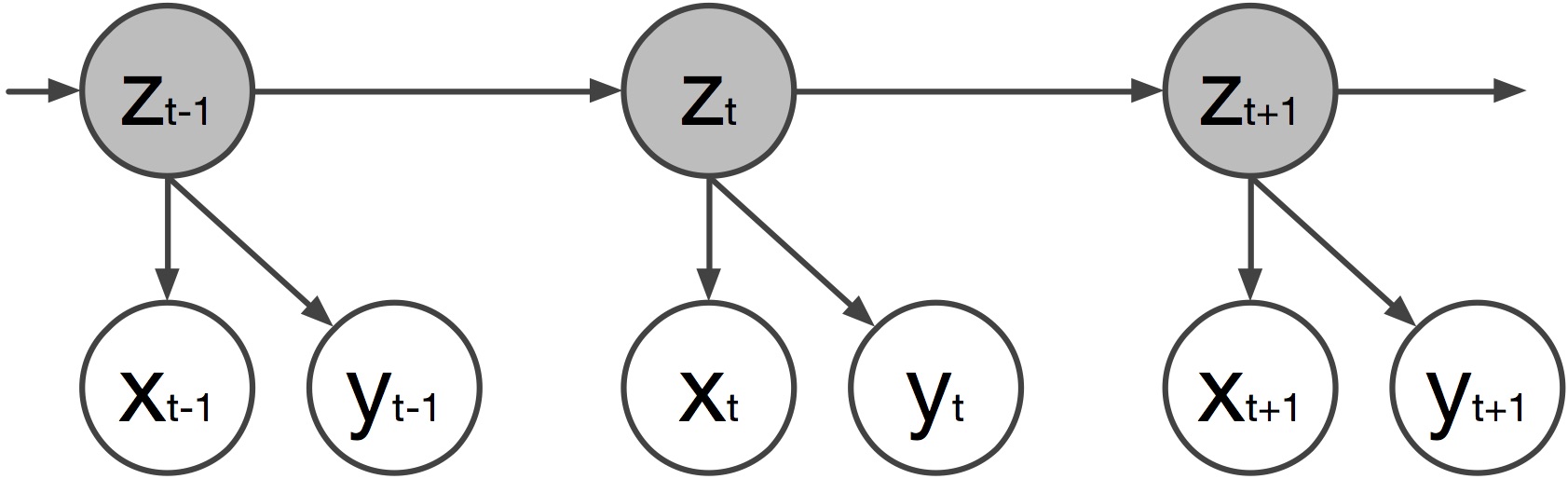}
\caption{The DBN representation of the multivariate HMM used in the presented method.}
\label{fig6}
\end{figure}

Given the two datasets, the IMU's corresponding feature vectors should be mapped to those of the full-body motion. The proposed method used a multivariate HMM. The basic advantage of such a mapping process is its ability to allow the prediction of missing features, especially when dealing with multivariate data. In the multivariate mapping, it is assumed that the same underlying process generates the full-body motion and the IMU data by jointly representing their observation sequences. Figure \ref{fig6} illustrates the dynamic Bayesian network (DBN) of the multivariate HMM process.

For the multivariate mapping process, the feature vectors at time $t$, constructed from the full-body motion ($x_t$) and the single IMU ($y_t$), are concatenated to form $z_t$ that denotes the $i-th$ state of the HMM at time $t$. This concatenation is represented as:

\begin{equation}
Z=[X,Y]=\begin{bmatrix}
x_{11} & \cdots & x_{1_{d_x}} & y_{11} & \cdots & y_{1_{d_y}}\\ 
\vdots & \ddots & \vdots & \vdots & \ddots & \vdots\\
x_{F_1} & \cdots & x_{F_{d_x}} & y_{F_1} & \cdots & y_{F_{d_y}} 
\end{bmatrix}
\label{eq1}
\end{equation}
where $X$ and $Y$ denotes the feature vector of the full-body motion sequence and the  feature vector of the IMU's data ($F$ sequences each) respectively, and $d_x$ and $d_y$ are the respective dimensions of the corresponding feature vectors. The probability distribution of the HMM at time $t$ of the $i-th$ state is defined as a joint multivariate Gaussian distribution:

\begin{equation}
p(x_t, y_t | z_t) = \mathcal{N} ([x_t, y_t]; \mu_i,U_i)
\label{eq2}
\end{equation}
where the concatenation of the mean of the distribution is defined by $\mu_i$ and expressed as follows:

\begin{equation}
\mu_i = [\mu_i^X, \mu_i^Y]
\label{eq3}
\end{equation}

$U_i$ is the covariance matrix that can be decomposed into four submatrices representing the univariate and crossvariate covariances between the full-body motion and IMU's data. The decomposition of the covariance matrix $U_i$ is represented as follows:

\begin{equation}
U_i=\begin{bmatrix}U_i^{XX} & U_i^{XY} \\ U_i^{YX} & U_i^{YY} \end{bmatrix}
\label{eq4}
\end{equation}
Given the datasets that should be mapped, the parameters of the multivariate HMM are estimated based on the widely used expectation maximization (EM) algorithm \cite{ref47}. 

In the presented method, the predictions of the joint observation distribution are converted to conditional distribution by marginalizing over the sensor's observation vectors at time $t$, using the following equation:

\begin{equation}
p( x_t | y_t, z_t ) = \mathcal{N} (x_t; \check{\mu}_i^X (y_t ), \check{U}_i^{XX})
\label{eq5}
\end{equation}
In Equation \ref{eq5}, the mean $\check{\mu}_i^X (y_t )$ and the covariance $\check{U}_i^{XX}$ of a virtual character's full-body locomotion is estimated by combining the mean and the covariance of the full-body motion with a linear regression over the sensor's features, as follows:

\begin{equation}
\check{\mu}_i^X (y_t ) = \mu_i^X + U_i^{XY} (U_i^{YY} )^{-1} (y_t - \mu_i^Y )
\label{eq6}
\end{equation}
and

\begin{equation}
\check{U}_i^{XX} = U_i^{XX} - U_i^{XY} (U_i^{YY})^{-1} U_i^{YX}
\label{eq7}
\end{equation}
Concluding, with the foregoing approach, the data provided by the IMU is efficiently mapped with the full-body captured motion.

\subsection{Hierarchical HMM}
\label{sec52}

\begin{figure}[htb]
\centering
\includegraphics[width=0.45\textwidth]{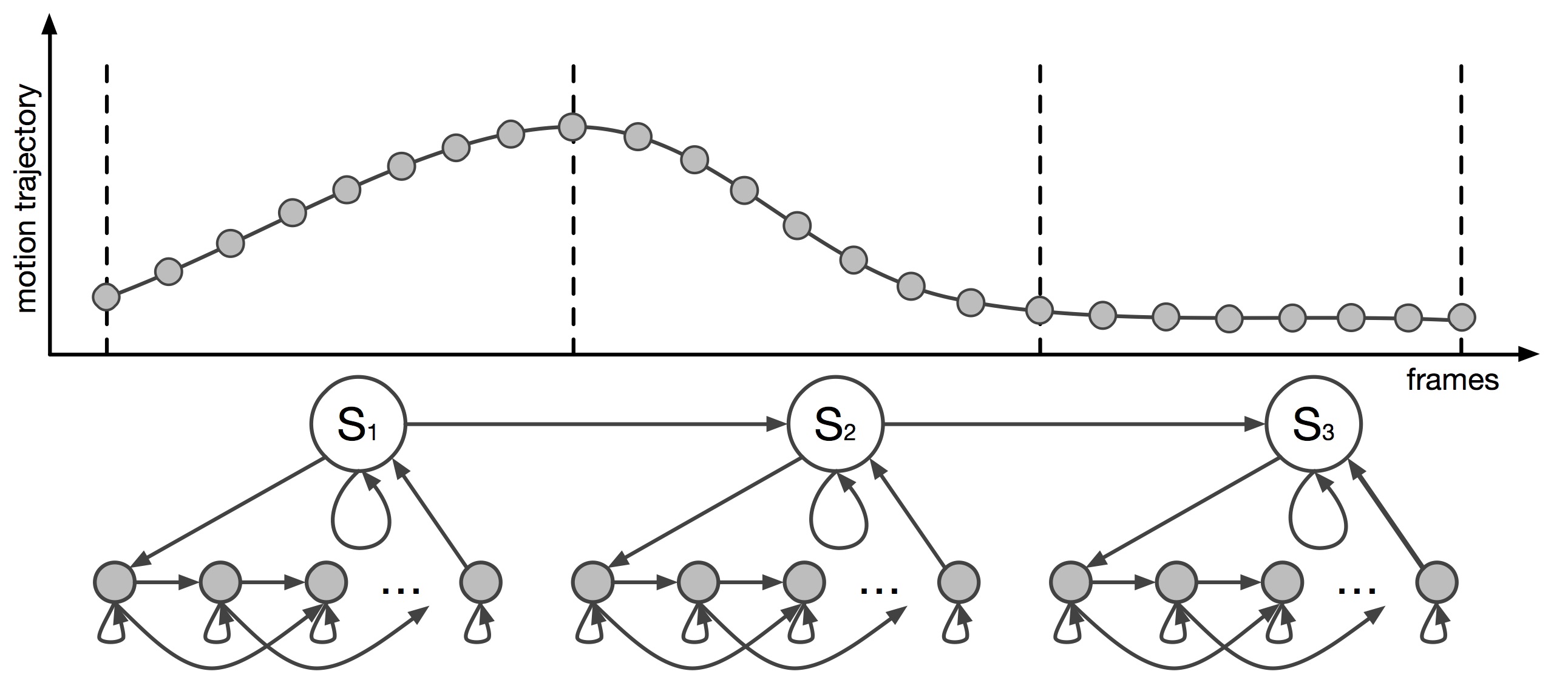}
\caption{The model is built from a sample trajectory.}
\label{fig7}
\end{figure}

The aforementioned multivariate HMM was extended to handle the motion data in a hierarchical way. The hierarchical HMM was implemented by adopting a template-based learning method (see Figure \ref{fig7}), which was used to build the model parameters from a single example. Each of the locomotion phases was assigned to the higher level of the model. Each of the higher levels of the model contained the motion's lower-level structure, which comprised the frame structure of the motion. The lower level of the proposed HMM was built by assigning a frame-based state to every frame of the motion segment with a structure that followed a left-to-right transition process. Given the user's input motion, the segment phase recognition and the motion reconstruction processes were performed in real-time, using the forward algorithm (see Section \ref{sec6}). Based on the input data, the forward algorithm was responsible for predicting the most probable segments, their associated weights according to the input feature vector, and the time alignment of the locomotion phase to the sample motion data. This approach allowed a character's full-body locomotion sequences to be reconstructed in real-time, using a single IMU, by combining the reconstructed segment phases from the different locomotion behaviors.

\subsubsection{Training Process}
\label{sec521}

Since the HMM was characterized by two different hierarchical levels, each responsible for a different part of the motion reconstruction process, the training process of the model was also separated into two steps. The first step trained the hierarchy's higher level that contained the segment phases of the locomotion. The second step trained the system to recognize the progression of the motion sequence and to reconstruct the motion. In the proposed HMM, both hierarchies of the model were encoded with a left-to-right transition structure, and a left-to-right Markov chain was used to define each learning example. The left-to-right transition structure did not include a backward transition, meaning that the model stayed in the same state, progressed to the upper state, or exited a state to search for the next probable one. This model was chosen because it accurately represented a temporal system. 

The first step of the training process trained the HMM's higher hierarchy, which contained the segment phases of the motion data. Three possible transitions (see Figure \ref{fig5}) between the segments were chosen (this three-way transition process allowed the model to freely reconstruct the motion sequences). Considering that the data is regularly sampled, the transition probabilities are set manually as follows:
\begin{itemize}
\item \textbf{self-transition:} $a_{self}^{phase}=1/3$,
\item \textbf{next transition:} $a_{next}^{phase}=1/3$, and
\item \textbf{exit transition:} $a_{exit}^{phase}=1/3$.
\end{itemize}

Self-transition ($a_{self}^{phase}$) meant that the input motion remained in a particular locomotion phase, which indicated that such a phase would still evolve. Next transition ($a_{next}^{phase}$) meant that the input motion would continue to the next locomotion phase, after having predicted the complete segment phase normally. Therefore, the locomotion behavior evolved normally. Finally, exit transition ($a_{exit}^{phase}$) meant that the input motion did not belong to the current segment. Thus, the system exited the segment state and searched for the next probable one from a different locomotion behavior. Note that the exit state is reached when the user decides to change the locomotion behavior that he or she performed in the middle of an evolved locomotion phase.

To train the model to recognize the locomotion phase where the input trajectory belonged, a temporal profile of the trajectory was built to create a left-to-right HMM by associating each segment state directly with an HMM state. Each of the HMM states corresponded to a sample in the training data and was associated with a Gaussian probability distribution $p( i | S)$, which was used to estimate the probability of an observation sequence $S$ for the segment state $i$ of the model and is represented as follows:

\begin{equation}
p(i | S) =\frac{1}{H}\sum_{h=1}^H \frac{1}{\sigma_{i,h} \sqrt{2\pi}}\exp\left[-\left( \frac{(S_h - \mu_{i,h})^2}{2\sigma_{i,h}^2}\right) \right]
\label{eq8}
\end{equation}
where $i$ denotes the $i-th$ sample associated with the segment state $S$, and $\sigma_{i,h}$ represents the standard deviation. Based on this equation, the model parameters of the HMM are refined by using the complete number of motion segments $H$ that belong to a particular segment phase of the locomotion. 

The second step trained the HMM's lower hierarchy, which contained the frames of the motion segment. The model was trained by using a method similar to the training process for the segment state. However, the system was not trained to recognize the motion segments but to recognize the motion's progress. As previously described, the frame steps of a motion segment were also encoded in a model represented by a left-to-right transition structure. Given a motion segment $Y$, its sequence was used to set a left-to-right Markov chain. As before, three possible transitions (see Figure \ref{fig5}) between the frames of the segment were chosen. Moreover, by also considering that the data is regularly sampled, the transition probabilities are set manually as follows:
\begin{itemize}
\item \textbf{self-transition:} $a_{self}^{frame}=1/3$,
\item \textbf{next transition:} $a_{next}^{frame}=1/3$, and
\item \textbf{skip transition:} $a_{skip}^{frame}=1/3$.
\end{itemize}

In this case, the self-transition ($a_{self}^{frame}$) meant that there was no progression of the motion (e.g., when the user remained motionless). The next transition ($a_{next}^{frame}$) indicated that the motion would evolve normally, therefore, the new reconstructed frame would be computed. The skip transition ($a_{skip}^{frame}$) was responsible for helping the model evolve in case the performer's input data stopped following the sample behavior. For example, users might consider changing their walking behavior to a running or jumping one before the progression of the motion would reach the exit frame. In this case, the $a_{skip}^{frame}$ transition would help the motion progress to exit the state and help the motion reconstruction process to adapt to the new motion quickly. The proposed method's ability to provide continuous smooth motion even when users decide to change their behavior is advantageous. Note that the $a_{skip}^{frame}$ is associated with the $a_{next}^{phase}$ transition of the segment phase since it helps the motion to evolve (faster but normal) before terminating the reconstruction process of a particular motion segment. This is quite important in the presented method, especially when the system loses the tracking of the sensor. In such a case, the reconstructed motion might suddenly jump, therefore, unnatural motion would appear. In such a scenario, the system would enter a segment phase, which would evolve normally (even if it was wrong). In case the system gained the track back, thanks to the $a_{exit}^{phase}$ the $a_{skip}^{frame}$ transition would help it reach the last frame of the segment fast enough. Then, the system would exit the phase normally and continue to the next predicted phase. Therefore, sudden changes in the resulting motion do not happen. The developed frame skip and phase exit abilities of the presented system conflict with previously proposed approaches \cite{ref45}\cite{ref52} that provide the ability to synthesize highly responsive, smooth and continuous motion. 

To train the presented model's lower hierarchical level, a temporal profile of the trajectory was built, which was later used to create a left-to-right HMM by associating each frame state directly with a state of the HMM to learn the motion progression phases. Each state corresponded to a sample in the training data and was associated with a Gaussian probability distribution function $p(i | Y)$, which was later used to estimate the probability of an observation sequence $Y$ for the frame state $i$ of the model. It is represented by the following equation:

\begin{equation}
p(i | Y) = \frac{1}{\sigma_i \sqrt{2\pi}}\exp\left[-\left( \frac{(Y - \mu_i)^2}{2\sigma_i^2}\right) \right]
\label{eq9}
\end{equation}

In Equation \ref{eq9}, $\mu_i$ denotes the $i-th$ sample associated with the frame state $i$, and $\sigma_i$ indicates the standard deviation. In the segment learning process, the model parameters were not refined by using the motion segments that belonged to the same phase. The system learned the individual parameters of the model, and the model parameters of each trajectory were then used in the proposed reactive interpolation functionality (see Section \ref{sec53}). Based on this training process, the two chosen states progressed to a motion, stayed motionless (remained in the current state), or exited as fast as possible. The exit state ability was not provided directly because discontinuities between motions appeared. Notably, in both cases (segment and motion progression phases), equal probabilities were chosen to be assigned at each transition state, which allowed the system to search any time step if the locomotion phase and the user's full-body motion closely resembled the output. This approach also allowed the user to perform different motions one after another and the system to reconstruct the required locomotion efficiently by combining the separately retrieved segments.

\subsection{Reactive Model Interpolation}
\label{sec53}

This reactive interpolation function allowed for interpolation between existing frames of the trajectories that were already in the database, which meant that the new reconstructed motion could more precisely follows the user's performance. Various interpolation methods can be used. Some are presented by Yoshimura et al. \cite{ref7}. Examples of such methods include interpolation between observations, interpolation between output distributions of HMM states that take state occupancies into account, and interpolation based on Kullback's information measure \cite{ref60}. This case adopted the first method (interpolation between observations). 

For the reactive model interpolation, the model parameters $\lambda_{t,1},...,\lambda_{t,D}$ that represent a $D$ number of motion segments $S_1,...,S_D$ that belong to the same segment state were defined. Moreover, $\bar{\lambda}_t$ was defined as a model of a trajectory $\bar{S}$ obtained by interpolating $K$ sample trajectories at time $t$. Based on the preceding explanation, a feature vector $y_t$ of a new motion segment $\bar{S}$ was obtained by linearly interpolating the observation vectors $y_{t,1},...,y_{t,D}$ of the motion variation, as follows:

\begin{equation}
\bar{y}_t = \sum_{k=1}^K a_k y_{t,k}
\label{eq10}
\end{equation}
where $\sum_{k=1}^K a_k \approx 1$. Now, given the mapped mean vector $\check{\mu}_i^X (y_t )$ and the mapped covariance matrix $\check{U}_i^{XX}$, it was possible to compute the new mean vector $\bar{\mu}$ and the new covariance $\bar{U}$ of the Gaussian output distribution $\mathcal{N}( \bar{y}, \bar{\mu}, \bar{U})$, as follows:

\begin{equation}
\bar{\mu} = \sum_{k=1}^K a_k \check{\mu}_{i,k}^X (y_t )
\label{eq11}
\end{equation}
\begin{equation}
\bar{U} = \sum_{k=1}^K a_k \check{U}_{i,k}^{XX}
\label{eq12}
\end{equation}

In this case, note that even if the model's parameters $\lambda_{t,k}$ (where $1 \leq k \leq K$) of the sample sensor's data had a tying structure, it was possible to obtain the model $\bar{\lambda}_t$ directly by interpolating $\lambda_{t,k}$. However, the model's parameters $\lambda_{t,k}$ had different structures from each other when the context clustering was independently performed for each motion variation model at the training stage. Therefore, it was difficult to obtain $\bar{\lambda}_t$ by interpolating $\lambda_{t,k}$, considering the model structure. To overcome this issue at the reconstruction stage, according to Yamagishi et al. \cite{ref48}, a number of $K$ sequences were engaged from $\lambda_{t,k}$ independently, and then a pdf sequence corresponding to $\bar{\lambda}_t$ was obtained by interpolating these $K$ pdf sequences. Next, a motion parameter sequence was generated from the interpolated pdf sequence. By means of linear interpolation, up to $K$ number of models could be used to obtain $\bar{S}$ and its corresponding $\bar{\lambda}_t$ model. Although in theory (according to \cite{ref41}\cite{ref48}), $K$ can grow up to $\infty$, the experimentation showed that $K=5$ provided reasonable results. Finally, it should be noted that an important reason to use this approach is that in the presented method, it is practically impossible and computational expensive to create in advance all possible $\bar{\lambda}_t$ models for every possible user input.

\section{Full-Body Locomotion Reconstruction}
\label{sec6}

With the sensor's input signal, the system was able to predict and reconstruct a virtual character's full-body motion during the application's runtime. This process was achieved by using the forward algorithm and considering the interpolated parameters of the model mentioned in Section \ref{sec53}. According to Rabiner \cite{ref6}, who provides a detailed explanation of the forward algorithm (see Appendix A as well), it is possible to efficiently predict the probability distribution of a sequence of observations based on the forward process.

In the proposed method, given the input signal from an IMU, $y_t$, the system was able to estimate the locomotion phase where the input motion belonged, predict the corresponding motion trajectories, and reconstruct them through interpolation. To achieve these steps, the forward inference procedure proposed by Murphy and Paskin \cite{ref8} was used. The forward inference procedure evaluated the most likely locomotion phase based on the user's input motion, and then the progress of the motion's trajectory. Within a recognized locomotion phase, the progression of a character's motion was represented by a left-to-right transition, which respected the motion segments' sequential order.

According to this explanation, with the input feature vectors $y_1,...,y_t$, the locomotion phase $S_h$ where the input motion belonged was estimated by computing the probability $p(S_h | y_1,..., y_t)$, using the forward algorithm. This procedure allowed the input data from the IMU (motion trajectory) to be segmented during the application's runtime in the sense that the most probable phases of the locomotion were updated in each time step. This process resembled the forward pass of the widely used Viterbi algorithm. Although each locomotion phase $S_h$ contained sequences of the substates that formed the full-body locomotion of a virtual character, using the summation (instead of the maximization of the model) of the model parameters made it possible to recognize the phase where the input locomotion belonged. 

In the next step, given the locomotion phase $S_h$ of the input motion, the virtual character's new full-body motion could be reconstructed. The full-body motion $\bar{x}_t$ was predicted by using the maximum likelihood function based on the state probabilities estimated with the forward algorithm as previously described. However, in this case, the presented method estimated $\bar{x}_t$, given $\bar{y}_1,...,\bar{y}_t$ ($\bar{y}_t$ denotes the results of the reactive interpolation functionality), $\bar{z}_t$ and $S_h$, as follows:

\begin{equation}
\bar{x}_t = \sum_{i} a_t (i, S_h)\arg\max_{x_t} \Big[p(x_t | \bar{y}_t, \bar{z}_t, S_h)\Big]
\label{eq13}
\end{equation}
where $a_t(i, S_h)$ defines the probability of the partial motion observation sequence at motion state $i$ and segment state $S_h$ at time $t$, taking into account the interpolated model parameters $\bar{\lambda}_t$, and is represented as follows:

\begin{equation}
a_t (i) = p(\bar{y}_1,..., \bar{y}_t, \bar{z}_t , S_h | \bar{\lambda}_t)
\label{eq14}
\end{equation}
Based on this process, the system reconstructed the full-body pose of a virtual character, given the signal provided by a single IMU.

\begin{figure*}[htb]
\centering
\includegraphics[width=1\textwidth]{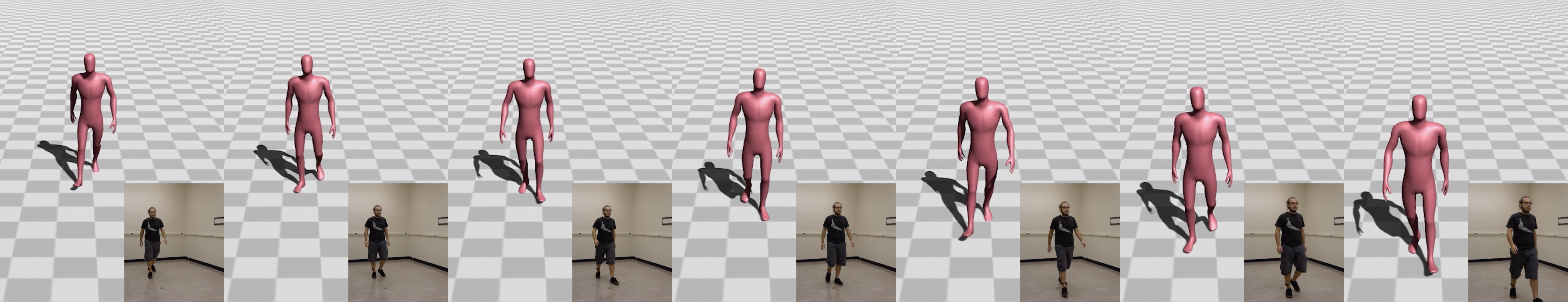}
\caption{Example of walking motion reconstructed with the proposed method.}
\label{fig8}
\end{figure*}

\begin{figure}[htb]
\centering
\includegraphics[width=1\columnwidth]{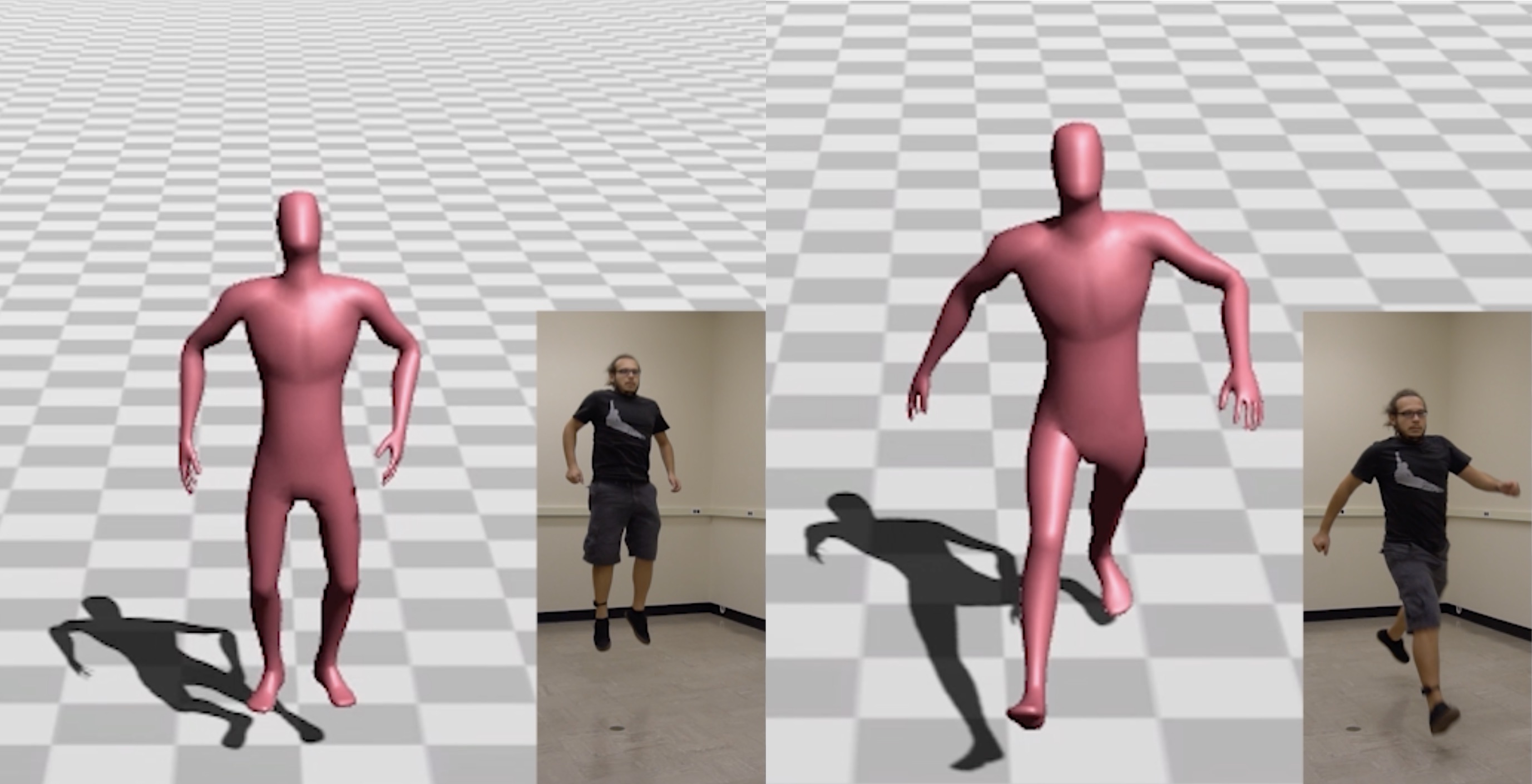}
\caption{Hopping and jumping poses reconstructed with the proposed method.}
\label{fig9}
\end{figure}

The segment phase recognition process of the presented model enables users to perform freeform actions. This means that the general pattern of locomotion might not be followed. Therefore, discontinuities between the reconstructed locomotion phases might appear during the reconstruction process. To avoid reconstructing motions that do not look natural enough, a velocity-based blending algorithm proposed by Levine et al. \cite{ref57} was used to interpolate the corresponding reconstructed motion segments. Moreover, to avoid foot sliding and ground penetration effects, a simple inverse kinematic technique was applied to remove such artifacts. Figures \ref{fig8} and \ref{fig9}, as well as this paper's accompanying video, show sample reconstructed poses.

\section{Implementation Details and Evaluations}
\label{sec7}

This section is divided into two parts. The first part presents a few details regarding the implementation of the proposed method. The second part discusses the conducted evaluations to provide an understanding of the efficiency of the proposed approach.

\subsection{Implementation}
\label{sec71}

The proposed reconstruction method was implemented by using the Perception Neuron \cite{ref9} system with an 18-neuron setup to capture the performer's full-body motion sequence, along with a 3-Space$^{TM}$ wireless IMU \cite{ref46} as the single sensor. Each neuron from the Perception Neuron and the 3-Space$^{TM}$ wireless IMU house a gyroscope, accelerometer, and magnetometer. The method was implemented with an Intel i7-6700 CPU at 3.4 GHz with an NVIDIA GeForce GTX 1060 with 6GB and 16GB of RAM. All presented results are based on the mentioned system. Finally, for the HMM, window sizes ranging from 5 to 50 ms were tested. A 20 ms window size worked for all the motion types that were used.

\subsection{Evaluations}
\label{sec72}
To evaluate the efficiency of the presented methodology, four different evaluation studies were conducted. First, the framerate and the latency of the motion reconstruction process were computed. Second, positioning the IMU on different body parts showed the sensor's optimal position and that the method worked. Third, the method was compared with different approaches that mentioned full-body motion reconstruction from either a single IMU or a reduced numbers of sensors. Finally, the transition between motion segments was also evaluated perceptually against two other motion synthesis methods.

\subsubsection{Framerate}
\label{sec721}

The framerate of the proposed solutions was computed for a variety of scenarios and of motion data. Table \ref{tab2} presents the framerate and the latency for different motion lengths (in frames). Latency denotes the time required by the algorithm to reconstruct the character's full-body pose, given the input data. The rendering time and the velocity-based blending algorithm are excluded from the computation of the latency. While the method worked in real-time in all cases, when the number of frames (and consequently, the number of motion segments) increased, the latency also increased, and the framerate decreased. Notably, even when 10000 frames were used, the method was able to reconstruct the character's motion at 12 frames per second. However, in cases where up to 5000 frames were used, the reconstruction process worked enough faster, reaching 38 frames per second. Based on the presented results, when an extensive number of sample motions are used, the methodology's framerate decreases. However, in scenarios that do not require a precise reconstruction of human locomotion but an affordable solution for reconstructing simple locomotion behaviors (e.g., walking and running behavior of a large group of pedestrians), the presented single IMU locomotion reconstruction method could be ideal.

\begin{table}
\centering
\caption{Results obtained when evaluating the performance of the presented methodology using different motions (W: walking, R: running, J: jumping, and H: hopping) and data sizes.}
\begin{tabular}{| p{1.3cm} | l | l | l | l |}
\hline
\textbf{Motion Type}	& \textbf{Frames} & \textbf{Segments} & \textbf{Framerate} & \textbf{Latency} \\
\hline
\hline
W			& 500		& 56			& 65 fps			& 0.025 s \\
\hline
W			& 1000		& 119		& 62 fps			& 0.039 s \\
\hline
W			& 2000		& 254		& 57 fps			& 0.073 s \\
\hline
W			& 4000		& 498		& 49 fps			& 0.124 s \\
\hline
W-R			& 5000		& 663		& 38 fps			& 0.292 s \\
\hline
W-R-J-H		&10000		& 1387		& 12 fps			& 0.758 s \\
\hline
\end{tabular}
\label{tab2}
\end{table}

\subsubsection{Evaluating IMU Positioning}
\label{sec722}

\begin{figure}[htb]
\centering
\includegraphics[width=1\columnwidth]{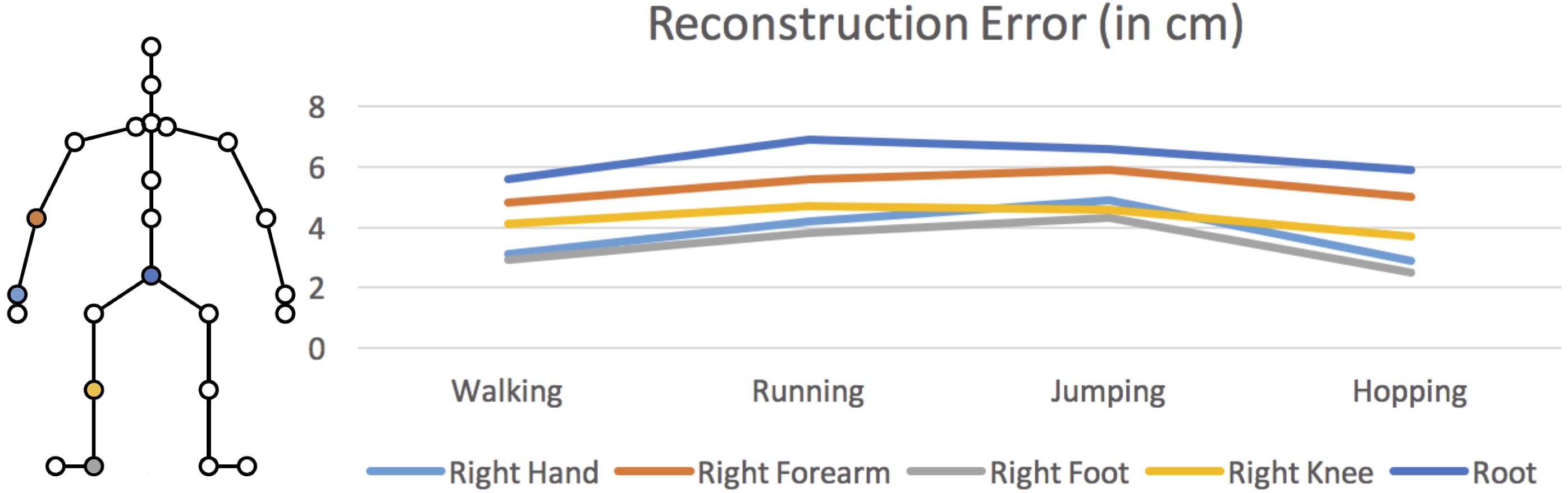}
\caption{The position of the IMU (left) and the reconstruction error when compared with the ground truth data (right).}
\label{fig10}
\end{figure}

Locomotion sequences using a single IMU placed on different body parts were reconstructed to evaluate the efficiency of the methodology. In this evaluation, the whole development process (from motion capture to HMM training) was iterated. Five different body parts were chosen: right hand, right forearm, right foot (which was initially captured), right knee, and root (see Figure \ref{fig10}). Moreover, to test the method's robustness, four different motions were reconstructed: walking, running, jumping, and hopping. The mean square error (MSE) approach was adopted to measure the reconstruction errors, which were computed between the positions of the joints of the virtual character, after excluding the global root position.

\begin{figure*}[htb]
\centering
\includegraphics[width=1\textwidth]{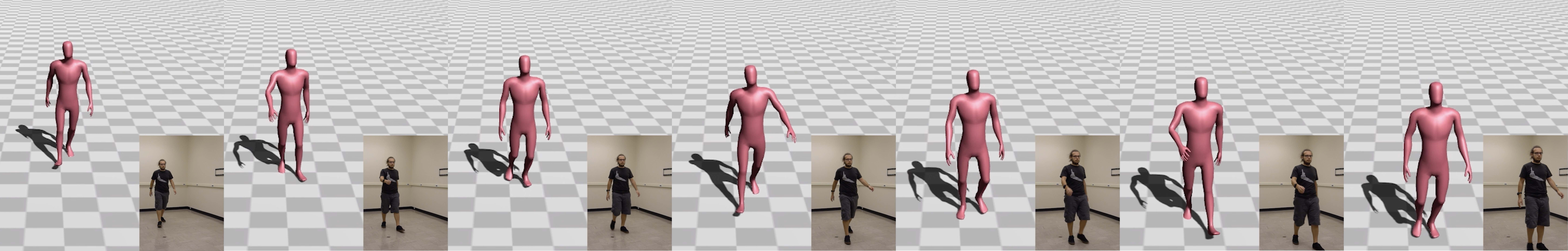}
\caption{Example of walking motion reconstructed with the proposed method when the IMU placed on user's right hand.}
\label{fig11}
\end{figure*}

Figure \ref{fig10} presents the findings of this evaluation process, which clearly show that the proposed method closely reconstructed the motions for each IMU position. However, the motion sequences that were reconstructed when the IMU was placed on the foot or the hand had less error compared with the forearm, the knee, or the root. When the IMU was placed on the root, the reconstruction error was greater than on other body parts. Finally, Figure \ref{fig11} shows a walking motion sequence reconstructed when placing the IMU on user's right hand.

\subsubsection{Reconstruction Evaluation Against Previous Approaches}
\label{sec723}

Previously, only Eom et al. \cite{ref12} had proposed a method of reconstructing a virtual character's full-body locomotion using a single sensor. Thus, their method was used as a basis for evaluating the reconstruction errors of this present study's method. The proposed approach was also compared against two methods that used two sensors, proposed by Riaz et al. \cite{ref10} and Min and Chai \cite{ref11}. As before, the reconstruction errors were computed between the positions of the joints of the virtual character.

During the evaluation process, four different motions (walking, running, jumping, and hopping) were used as ground truth data. For all the conducted evaluations, the MSE was computed with the ground truth data. The results obtained (see Table \ref{tab3}) showed that the proposed solution had fewer errors than those of Eom et al.'s \cite{ref12} method for all four motion types. However, compared with the methods of Riaz et al. \cite{ref10} and Min and Chai \cite{ref11}, the proposed solution provided inadequate results. Because the two previous studies used two sensors, their process for reconstructing a virtual character's locomotion was more constrained. For a fairer comparison with the two-sensor solutions, the proposed method's ability to reconstruct a virtual character's full-body locomotion by using two IMUs (hand and foot) was also tested. The process was iterated, and the MSE was recomputed. The results obtained from this additional evaluation (Table \ref{tab3}) clearly indicate that the presented method can reconstruct motion sequences with fewer errors when signals from two IMUs are used. This demonstrates the proposed statistical model as a powerful solution for reconstructing a variety of locomotion behaviors of virtual characters in real-time.

\begin{table*}[htb]
\centering
\caption{Results obtained when evaluating the presented method with previously proposed solutions. The reconstruction error is measured in cm.}
\begin{tabular}{| l | l | p{1.5cm} | p{1.6cm} | l | l | l |}
\hline
\textbf{Motion Type}		&\textbf{Frames}	&\textbf{Proposed (1 sensor)}	& \textbf{Proposed (2 sensors)} & \textbf{Eom et al.} \cite{ref12} & \textbf{Riaz et al.} \cite{ref10} & \textbf{Min and Chai} \cite{ref11} \\ 
\hline
\hline
Walking				& 2000		& 2.74		& 2.07		& 3.54		& 2.53		& 2.26		\\
\hline
Running				& 2000		& 2.79		& 2.11		& 4.15		& 2.71		& 2.35		\\
\hline
Jumping				& 1000		& 4.17		& 1.98		& 4.98		& 3.97		& 3.62		\\
\hline
Hopping				& 1000		& 2.46		& 1.91		& 4.42		& 2.44		& 2.19		\\
\hline
\textbf{Total Dataset}		& 6000		& 4.53		& 3.03		& 7.33		& 4.03		& 3.77		\\
\hline
\end{tabular}
\label{tab3}
\end{table*}

\subsubsection{Evaluating Transition Between Motion Segments}
\label{sec724}
To understand the presented methodology's efficiency in synthesizing natural-looking motion sequences, a perceptual evaluation study examined the transition between two motion segments. The proposed method was evaluated against Min and Chai's \cite{ref11} solution in which the transition was learned by analyzing the motion primitives and was performed by using a statistical approach, as well as against the motion graph approach proposed by Kovar et al. \cite{ref58}. Particularly, especially the evaluation against motion graphs (Kovar et al. \cite{ref58}) could provide good insights into the efficiency of synthesizing natural-looking motions.

\begin{figure}[htb]
\centering
\includegraphics[width=1\columnwidth]{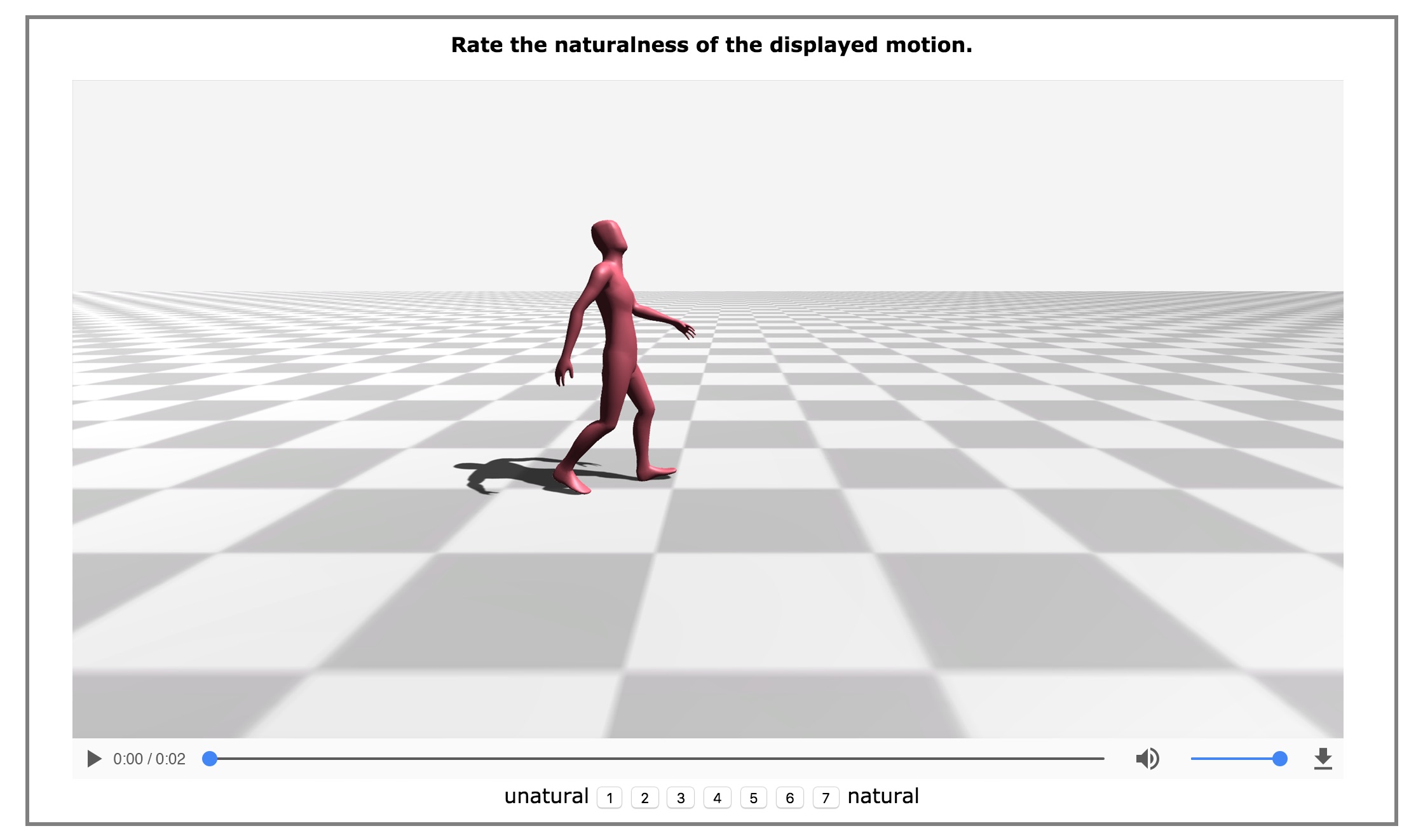}
\caption{A screenshot from the conducted experiment.}
\label{fig12}
\end{figure}

In this evaluation, five transitions were considered: from idle to walking motion, from walking motion to idle, from walking to walking motion, from walking to running motion, and from running to walking motion. For each motion type, eight random motion segments were considered. Therefore, given the five transitions, for each of the examined methods (the proposed one and those of Min and Chai \cite{ref11} and Kovar et al. \cite{ref58}), 40 motions were synthesized (120 in total). The videos captured by placing the virtual camera on the character's right side. Therefore, the subjects could look properly at each of the synthesized transitions. In each iteration of the experiment, the subjects watched a single video. The animations were captured in such a way that the transition started at the end of the first second. None of the videos exceeded the two-second duration. The videos were displayed in a random order, and the subjects were asked to rate the naturalness of the synthesized motion transition on a seven-point Likert scale (from 1 = unnatural to 7 = natural). Figure \ref{fig12} shows a screenshot of the developed experiment. In total 25 participants (19 males and 6 females, aged 21-28) attended the perceptual evaluation study. The evaluation study took no more than 20 minutes. It is noteworthy that all the participants completed it.

The results obtained from this experiment were analyzed by using paired t-tests. Evaluating the proposed method against those of Min and Chai \cite{ref11} ($t(24) = 0.047$, $p = 0.078$) and Kovar et al. \cite{ref58} ($t(24) = 0.064$, $p = 0.091$) showed no significant difference in the results ($p > 0.05$ in both evaluations) in terms of naturalness of the synthesized transition between motion segments. The obtained results indicate that even when synthesizing the transitions by using a simple velocity-based, motion-blending algorithm, the synthesized motion, and consequently the transition between the motion segments, can be characterized as perceptually consistent.

\section{Conclusions}
\label{sec8}

This paper has presented a solution for reconstructing full-body locomotion sequences of virtual characters in real-time, using a single IMU. To overcome the problem's complexity, a method based on the construction of an HMM was used. In the proposed approach, a hierarchical multivariate HMM with reactive interpolation functionality was constructed wherein in the higher hierarchical level the locomotion phases were assigned, and in the lower hierarchical level the frame structure of the motion was assigned. The full-body motion reconstruction was achieved by using the forward algorithm. Evaluating the process showed it as ideal for real-time applications and highly efficient because it could reconstruct full-body locomotion by minimizing errors compared with previously proposed solutions.

Using a single sensor to reconstruct human locomotion is challenging. Even if it is possible to constrain a statistical motion model that describes human motion and reconstructs a character's full-body motion by minimizing errors, humans daily perform numerous motions and actions that cannot be recognized and reconstructed (e.g., upper-body motions, such as hand waving or punching, or lower-body motions, such as kicking with a foot that has no attached sensor). Therefore, the proposed solution's use is limited to cases requiring reconstruction of simple motions. Additionally, similar to other data-driven motion reconstruction methods, it is not possible to reconstruct motion sequences that are not included in the dataset. Working with a large dataset, as well as with a variety of motions and human actions, also increases the likelihood of ambiguities or difficulties (e.g., unnatural poses). In conclusion, the proposed solution is most appropriate for applications such as navigation in a virtual environment (where small divergences are inconsequential) or locomotion capture for crowd analysis and simulation-related applications.

Besides the advances presented in this paper, several issues should also be discussed. Among others, the chosen way of handling the motion data affects the computational time required for reconstructing a character's full-body motion. This point becomes quite clear when reviewing the obtained results. When the number of motion segments increases, the framerate decreases dramatically. The reason is that this type of method requires saving a large amount of data, and searching through all of it decreases the framerate. Therefore, solutions that learn general representation from motion examples (e.g., \cite{ref55}\cite{ref56}) should also be considered and tested in the future for reconstructing full-body motion sequences in such under-constrained scenarios.

\section*{Appendix: The Forward Procedure}
\label{appendix}

The forward algorithm (see Rabiner \cite{ref6}) can be efficiently used to predict the probability distribution of an output sequence $x_1,...,x_t$, given a sequence of observations $y_1,...,y_t$. This step is achieved by defining a forward variable, which is quite useful in avoiding numerical errors. In the presented method, for a predicted segment phase $S_h$, the forward variable (probability distribution variable) of the partial observation sequence until time $t$ and state $i$ is defined as $a_t (i, S_h) = p(\bar{y}_1,...,\bar{y}_1, \bar{z}_t, S_h | \bar{\lambda}_t)$. This process is computed inductively, based on an initialization and induction. Specifically, the initialization process computes $a_1(i, S_h) = \pi_i b_i(y_1)$, subject to $i \in [1,N]$, where $\pi_i$ denotes the initial distribution of state $i$ at the segment phase $S_h$, and $b_i(y_1)$ represents the observation probability distribution. The induction phase, it is computed as:

\begin{equation*}
a_{t+1} (i,S_h ) = \Bigg[ \underbrace{\sum_{i=1} a_t (i,S_h)a_{ij}}_\text{prediction}\Bigg] \underbrace{b_i (y_{t+1})}_\text{update}
\end{equation*}
subject to $t \in [1, T-1]$ and $i \in [1,N]$. Note that $a_{ij}$ signifies the transition probability distribution between state $i$ and $j$ of the HMM.

\bibliographystyle{ieeetr}
\bibliography{template}

\end{document}